\documentclass[aps,pre,onecolumn,floatfix]{revtex4-2}

\usepackage{epsf}
\usepackage{subfigure}
\usepackage{amsmath}
\usepackage{amssymb}
\usepackage{bm}
\usepackage{graphicx}
\usepackage{epstopdf}
\usepackage{float}
\newcommand{\mean}[1]{\left \langle #1 \right \rangle}

\newcommand{\be}{\begin{equation}}
\newcommand{\ee}{\end{equation}}
\newcommand{\bea}{\begin{eqnarray}}
\newcommand{\eea}{\end{eqnarray}}

\newcommand{\parent}[1]{\left( #1 \right)}

\begin{document}

\title{Control Strategies for Maintaining Transport Symmetries Far From Equilibrium}

\author{David Andrieux}

\noaffiliation

\begin{abstract}
We present two strategies for controlling the transport dynamics of mesoscopic devices. 
In both cases, we manipulate the system's output — such as particle currents and energy flows — while maintaining symmetric transport properties, even under far-from-equilibrium conditions. 
We provide exact descriptions of each scheme and investigate their characteristics. 
Notably, one of them minimizes the dissimilarity between the original and modified processes, as quantified by the Kullback-Leibler divergence.
These findings can be used to improve the design and control of mesoscopic systems.
\end{abstract}

\maketitle

\vskip 0,1 cm

\section*{Context and objectives}

Imagine you are operating a mesoscopic device designed to produce a specific output, such as a flow of particles or energy currents. 
The device's operating mode reflects a balance between achieving the desired transport properties and managing acceptable levels of fluctuations, dissipation, and robustness. 
Now, consider the problem of adjusting this output in a controlled way. 
How can you ensure that the device continues to function as intended?

In many systems, altering any single kinetic parameter can impact the overall dynamics, including fluctuations, dissipation, and efficiency \cite{H05}. 
Conversely, some parameters may exhibit "sloppy" behavior, meaning that altering their values over a wide range may not significantly change the output \cite{GEA07, MEA13}.
Navigating this complex parameter space to identify the parameters' adjustments for achieving the desired output is a challenging task. 

In addition, different transport processes are interconnected: changes to one current can affect others. 
Close to equilibrium, the Onsager symmetries describe how the response of a system is related to changes of reciprocal processes \cite{O31, GM11, NP77}.
This is particularly useful in materials science, electrochemistry, and biological processes, where the Onsager reciprocity allows predictions about one type of response based on measurements of its reciprocal process.
Another benefit of the reciprocity relations is the reduction in the number of independent coefficients needed to describe the system.

Extending these symmetries to far-from-equilibrium situations would expand our understanding of energy and matter transfer in various applications. 
In addition, predictable symmetry principles that relate one response to another even in the nonlinear regime would improve our ability to design and control mesoscopic devices that rely on multiple coupled processes (e.g., thermal, electrical, mechanical). For example, if a large applied temperature gradient not only generates a heat current but also influences mechanical deformation in a predictable way, designers could exploit these relationships to optimize multiple performance metrics simultaneously \cite{V18}.

However, extending the Onsager symmetries to the nonlinear regime has remained elusive \cite{P67}. 
A notable exception is the fluctuation theorem for currents, which imposes linear relations between the response coefficients of the same order \cite{AG04}. These constraints reduce the number of independent coefficients \cite{BG18}, but do not provide full symmetries. 

In this paper, we propose two control schemes designed to achieve a system's desired output together with symmetric transport properties. 
Our strategy is to decompose the space of parameters in terms of thermodynamics quantities and use geometric structures to navigate the space. 
This approach leads to the development of two distinct control schemes, referred to as $e$ and $m$, which are grounded in concepts from information geometry.

We obtain explicit constructions for both strategies. 
The $e$ dynamics coincide with the concept of "dynamical equivalence classes" introduced to study large current fluctuations \cite{A12c, A24c}. 
We also demonstrate that the $e$ dynamics modify the device's outputs while remaining as close as possible to the original dynamics, as measured by the Kullback-Leibler divergence.
In turn, the $m$ dynamics define new equivalence classes not considered previously.
We derive expressions for the affinity-current manifold and response coefficients for these classes, and express them in terms of equilibrium quantities.

These results enable the control of mesoscopic systems by providing the necessary combinations of parameters to achieve the desired outputs, and we illustrate this approach on two systems of interest. 
These findings also reveal hidden structures and symmetries within the parameter space with a clear thermodynamic interpretation. 
Beyond its potential applications, this work integrates and extends recent findings in information geometry, transport fluctuations, response theory, and the generation of Markov chains. 
\\

\newpage

\noindent {\bf Structure of the paper:}\\ 

In Section 1, we introduce the dynamics of Markov chains and their associated transport properties. We explore the relationship between microscopic and macroscopic quantities, as well as the dimensionality of the parameter space. 
We also present the framework for nonlinear response theory.

In Sections 2 and 3, we construct the $e$ and $m$ control strategies and analyze their properties. 
We demonstrate that both schemes lead to symmetric transport properties, albeit through different physical mechanisms.

In Section 4, we apply these concepts to two distinct systems. 
The first system is a molecular motor toy model with a single current, and which can be solved analytically. 
The second system involves two coupled currents and a higher-dimensional parameter space. 
Numerical simulations illustrate the nonequilibrium symmetries present in this more complex system.

Finally, in Section 5 we discuss the implications and new perspectives offered by these results.\\

        







\section{Transport in mesoscopic systems}

\subsection{Markov chain dynamics}

The dynamical properties of many biological, chemical and physical mesoscale systems are well-described by a Markov dynamics \cite{NP77, H05}. 
We thus consider the space $\Lambda$ of primitive Markov chains on a finite state space of size $N$ (an extension of our results to continuous-time is straightforward).
Each primitive chain $P \in \Lambda$ then admits a unique stationary state $\pi$ with $\sum_i \pi_i =1$.

Steady-state transport is characterized by the stationary currents
\bea
J_{ij} = \pi_i P_{ij} - \pi_j P_{ji} \, .
\label{Jij}
\eea
The system is at thermodynamic equilibrium when all currents vanish, i.e. when  
\bea
\pi_i P_{ij} = \pi_j P_{ji} \quad {\rm (detailed \, balance)}
\eea
for all pairs $(i,j)$.

The currents $J_{ij}$ are not independent as they must satisfy Krichoff's law at each node
\bea
\sum_{j} J_{ij} = \sum_{i} J_{ij} = 0 \, .
\label{Kirch}
\eea
As a result, there exist a set of {\it independent currents} $J_\alpha$ such that any current $J_{ij}$ can be expressed as a linear combination of these independent currents. 
These independent currents are linked to the fundamental cycles $c_\alpha$ of the graph associated with $P$ \cite{S76}:
\bea
J_{ij} = \sum_\alpha \gamma_{ij}(c_\alpha) J_\alpha \, .
\eea
Here the quantities $\gamma_{ij} (c)$ take the value $\pm 1$ if the transition $i \rightarrow j$ belongs to the cycle $c$ in the positive $(+)$ or negative $(-)$ direction, and $0$ otherwise.
An introduction to the graph representation of Markov chains and how to identify the independent currents is presented in Appendix A.

In tandem with the currents, the thermodynamic affinities driving the system out of equilibrium can be measured by the breaking of detailed balance along cyclic paths $c = (i_1, i_2, \cdots, i_n)$ as
\bea
A_c = \ln \parent{ \frac{P_{i_1i_2}\cdots P_{i_ni_1}}{P_{i_1i_2}\cdots P_{i_ni_1}} } \, .
\eea
The affinities vanish at equilibrium where detailed balance is satisfied. 
They do not involve the stationary probabilities $\pi$ and can thus be calculated from the transition probabilities directly.
Similar to the currents, only a subset $A_\alpha$ of these affinities are independent. 
The affinity of an arbitrary cycle can be expressed as a linear combination of the fundamental affinities (Appendix A).


The currents and affinities capture the anti-symmetric component of the dynamics. 
The symmetric component is described by the activities 
\bea
Y_{ij} = \pi_i P_{ij} + \pi_j P_{ji} \, .
\label{Yij}
\eea
These quantities (also called frenesies in the litterature) are not all independent since $\sum_{ij} Y_{ij} =2$.
Here, the analogue of the affinities are the "kinetic forces" 
\bea
\bar{X}_{ij} = \ln (P_{ij}P_{ji}) \, .
\eea
Eliminating one of the dependent activities $Y_r$ leads to effective forces \cite{A24c}
\bea
X_{ij} = \bar{X}_{ij} - \bar{X}_r \, .
\label{eff.force}
\eea
While less front-and-center than the currents and affinities, these symmetric components play an important role in shaping transport properties \cite{MW06, A24c}. 

\subsection{Minkowski structure of the parameter space}

In the previous section, we characterized the behavior of a mesoscopic system using macroscopic quantities like currents and affinities, linking them to the higher-dimensional space of microscopic parameters. 
In this section we analyze the dimensionality of the parameter space. 
Understanding the geometry of the parameter space and its relationship to thermodynamic quantities is crucial for apprehending - and ultimately controlling - the behavior of a system. 

A transition matrix $P$ can be represented by a graph with $N$ states, $S$ self-transitions, and $E$ edges, where each edge represents an allowed transition between states \cite{S76}.
The space $\Lambda$ of chains with a similar graph structure has dimension \cite{S76}
\bea
\text{dim} \, \Lambda =  2E+S-N  \, .
\eea
Since there are $E-N+1$ independent currents or affinities \cite{S76}, 
the iso-current manifold $\Sigma_{\pmb{J}}$ has dimension 
\bea
\text{dim} \, \Sigma_{\pmb{J}} = E+S-1 \, .
\eea
In particular, this holds for the equilibrium manifold $\Sigma_0 = \Sigma_{\pmb{J}=0}$.
Similarly, the iso-affinities manifolds $\Sigma_{\pmb{A}}$ have the same dimension $E+S-1$.\\

\noindent {\it Disordered ring example:} Consider a ring with $N$ states, $E = N$ edges and no self-transitions ($S=0$). 
There are $2N-N = N$ independent microscopic parameters and $N-N+1 = 1$ independent current and affinity. 
As a result, the equilibrium manifold, as well as all the iso-current manifolds, have dimension $N-1$.
The iso-affinity manifolds also have dimension $N-1$. 
This implies that one can change a combination of $N-1$ parameters while keeping the affinity or the mean current unchanged!\\

The dimensionality of the iso-currents and is-affinity manifolds reflects the non-unique, many-to-one mapping of microscopic parameters to thermodynamic (macroscopic) quantities.
This observation is reminiscent of sloppy models, where the behavior of the system is invariant despite variations in the microscopic parameters \cite{MEA13, GEA07}.
In terms of control, this indicates that there are multiple ways to navigate the parameter space to generate a given output. 
It also reveals that additional conditions are required to obtain a well-defined response theory. 
This last point will be detailed in the next sections.  

The space $\Lambda$ can be decomposed in terms the currents $J_\alpha$ and affinities $A_\alpha$ as well as the activities $Y_e$ and kinetic forces $X_e$.
Together with an appropriate divergence measure, these elements form an embedding of the parameter space with a Minkowski structure \cite{A24c, TQS20}.
Specifically, we have that the separation between two models $P$ and $P'$ take the form 
\bea
D_S ( P , P') &=& \frac{1}{2}  \sum_{\alpha} (A_\alpha-A_\alpha') (J_\alpha-J_\alpha') + \frac{1}{2}\sum_{e} (X_e -  X_e') (Y_e-Y_e')  \, ,
\label{DSKLb}
\eea
where $D_S$ is the symmetrized Kullback-Leibler distance. 

In this decomposition, the light cone for the $e$ variables correspond to the isokinetic and isoactivity manifolds. Remarkably, these light cones will provide us with the two control strategies we will use: the isokinetic manifold (constant $X_e$) corresponds to the $e$ control strategy while the isoactivity manifold (constant $Y_e$) corresponds to the $m$ strategy.

\subsection{Nonequilibrium fluctuations and response theory}

To study the behavior of a system, a typical quantity of interest is the system's current-affinity relationship $J_\alpha (\pmb{A})$, which expresses how the currents vary as a function of the driving forces or affinities.
However, since there are many more microscopic parameters than macroscopic affinities, any such function implicitly assumes a specific underlying trajectory in the parameter space. 

In the linear regime, the response coefficients only depend on the starting equilibrium, but not on the direction in parameter space. However, in the nonlinear regime this is not longer the case (Appendix B).
Put differently, the current-affinity manifolds are not uniquely defined unless all the microscopic parameters are specified \cite{H05, A23}. 

This is an important and subtle point, with theoretical and practical implications. 
As will we see below, both the $e$ and $m$ equivalence classes generate well-defined relationships $J_\alpha (\pmb{A})$ between currents and affinities by prescribing the values of all microcopic parameters.
The freedom to choose how to vary the microscopic parameters will also allow us to generate additional symmetries out of equilibrium.

Assuming for now that $J_\alpha (\pmb{A})$ is well defined in the above sense, we can study the behavior of the system as the affinities are varied.
The response of the currents is defined as
\bea
R_{\alpha \beta} (\pmb{A}) =  \frac{\partial J_\alpha}{\partial A_\beta} (\pmb{A}) \, .
\label{cur.resp}
\eea
The responses $R_{\alpha \beta}(\pmb{A})$ are symmetric at equilibrium, $R_{\alpha \beta}(\pmb{0}) = R_{\beta \alpha}(\pmb{0})$ \cite{O31}. However, this symmetry doesn't hold outside equilibrium in general.\\

\noindent {\it Note:} Some investigators instead introduce affinity-dependent responses $\sigma_{\alpha\beta}$ defined as \cite{T64, V18, V19}
\bea
J_\alpha = \sum_\beta \sigma_{\alpha\beta}(\pmb{A}) A_\beta \, .
\eea
However, as noted by Peterson \cite{P67}, in the nonlinear region $J_\alpha/A_\beta$ does not have the physical relevance that $\partial J_\alpha/\partial A_\beta$ has, although $\sigma$ has an intuitive appeal because of our familiarity with the linear form.\\

To further describe the response outside equilibrium, it is often convenient to expand the currents as functions of the affinities around $\pmb{A} =0$:
\bea
J_\alpha (\pmb{A}) = \sum_{l=1}^{\infty} \frac{1}{l!} L_{\alpha, \beta_1 ... \beta_l} \, A_{\beta_1} \cdot\cdot\cdot \ A_{\beta_l} \, ,
\label{J.exp}
\eea
where we sum over repeated indices. The response coefficients $L_{\alpha, \beta_1 ... \beta_l}$ describe the response of the system increasingly far from equilibrium. 
Equivalently, the affinities can be expressed as functions of the currents, $A_\alpha (\pmb{J})$.
Expanding around $\pmb{J} =0$ leads to the series expansion
\bea
A_\alpha (\pmb{J}) = \sum_{l=1}^{\infty} \frac{1}{l!} M_{\alpha, \beta_1 ... \beta_l} \, J_{\beta_1} \cdot\cdot\cdot \ J_{\beta_l} \, .
\label{A.exp}
\eea
We will use both formulations in our analysis. The expansion (\ref{J.exp}) will appear naturally when studying the $e$ equivalence classes while expression (\ref{A.exp}) will appear in the $m$ equivalence classes. 

The current fluctuations also provide important information about the systems' behavior. 
They are characterized by the generating function
\bea
G(\pmb{\eta}) =\lim_{n \rightarrow \infty} \frac{1}{n} \ln \mean{ \exp \parent{\pmb{\eta} \cdot \sum_{k=1}^n \pmb{J}(k)} }  \, .
\label{CGF}
\eea
The cumulants $K_{\alpha_1 ... \alpha_m}$ can then be obtained by successive derivations of $Q$ with respect to the counting parameters $\pmb{\eta}$.
In particular, the first-order cumulant $K_{\alpha} = J_\alpha$ while the higher-order cumulants capture higher-order statistics.

The cumulants $K_{\alpha_1 ... \alpha_m}$ also vary with the affinities.
Expanding the cumulants around $\pmb{A}=0$ leads to 
\bea
K_{\alpha_1 ... \alpha_m} (\pmb{A}) = \sum_{l=1}^{\infty} \frac{1}{l!} L_{\alpha_1 ... \alpha_m, \beta_1 ... \beta_l} \, A_{\beta_1} \cdot\cdot\cdot \ A_{\beta_l} \, .
\eea
The coefficients $L_{\alpha_1 ... \alpha_m, \beta_1 ... \beta_l}$ are not independent as the fluctuation theorem imposes linear relationships between coefficients of the same order $m+l$.
These constraints generate the full (Onsager) symmetries at first order and partial symmetries at higher orders \cite{AG04, BG18}.

In the next sections, we will describe our strategies to change the microscopic parameters of the system to control its output while ensuring that the system's nonlinear response remains symmetric away from equilibrium.


\section{e-control scheme}
\label{Sec.e.classes}

\subsection{Affinity-based representation of Markov chains}

To effectively control a Markov system or to generate a Markov chain with specific properties requires a parametrization of the space of transition matrices. 
Here we use a representation that generates Markov chains with given affinities. 
It will be useful to analyze the $e$ equivalence classes. 

From an equilibrium chain $H$, we can generate a Markov chain $P$ with affinities $A_\alpha$ through the mapping \cite{A12}
\bea
P = s[H \circ Z(\pmb{A})] \, .
\label{e.rep}
\eea
Here $\circ$ denotes the Hadamard product $(P \circ Q)_{ij} = P_{ij}Q_{ij}$ and 
\bea
Z_{ij} (\pmb{A}) \equiv \begin{cases} \exp \parent{\pm A_\alpha/2}  & \text{if $i\rightarrow j$ is a fundamental chord $\alpha$ in the positive (negative) orientation,} \\
                1 & \text{otherwise}.
                \end{cases}
\label{Z}
\eea
The mapping $s$ transforms a positive matrix $Q$ into a stochastic one as 
\bea
s[Q] = \frac{1}{\rho} \, {\rm diag}(x)^{-1} \, Q \, {\rm diag}(x) \, ,
\eea 
where $\rho$ is the largest eigenvalue of $Q$ and $x$ its corresponding right eigenvector \cite{FN01}.

Conversely, an equilibrium chain can be generated from $P$ through the mapping 
\bea
P^e = s\Big[ \parent{ P \circ P^{*} }^{(1/2)} \Big] \, ,
\label{Pe}
\eea
where $P^*$ is the time-reversed chain 
\bea
P^*_{ij} = P_{ji}(\pi_j/\pi_i)
\eea 
and $Q^{(1/2)}$ is the element-wise exponentiation. 
The chain $P^e$ is the equilibrium chain that has the closest projection distance to $P$ in the following information-theoric sense \cite{WW21}:
\bea
P^e = \arg \min_{Q \in \Sigma_0} D(Q|P) \, ,
\label{Pe.inf}
\eea
where the minimization is done over the space $\Sigma_0$ of compatible equilibrium chains \cite{FN02}. 
This hints at the central role that $P^e$ will play in the characterization of the $e$ equivalence classes.
The chain $P^e$ is also related to the entropy production of the system \cite{A24b}.

\subsection{Definition of the $e$ control scheme}

We are now in position to define our $e$ control scheme.
Starting from the reference dynamics $P$, the {\bf $e$-control scheme} is defined as
\bea
Q^e (\pmb{\lambda}) = s[P \circ Z(\pmb{\lambda})]  
\, .
\label{e.sol}
\eea
Here the parameters $\pmb{\lambda}$ directly control the system's affinities: If the chain $P$ has affinity $\pmb{A}$, the chain $Q^e (\pmb{\lambda})$ has affinities $\pmb{A}+\pmb{\lambda}$. 
The set of dynamics $Q^e$ forms the {\it $e$ equivalence classes}. These classes were initially introduced to study the large current fluctuations \cite{A12, A12c, V22}.\\

The class $Q^e$ forms a manifold with dimension equals to the number of independent currents or affinities.
The chains $Q^e (\pmb{\lambda})$ share similar characteristics for the symmetric part of their dynamics.
Indeed, noting that $Z_{ij}Z_{ji} = 1$, we see that two dynamics $P$ and $Q$ in the same class satisfy
\bea
P_{ij}P_{ji} \propto Q_{ij} Q_{ji} \, .
\label{e.def.alt}
\eea 
The conditions (\ref{e.def.alt}) imply that the kinetic forces are shifted by a constant along an $e$ equivalence classes, $\ln P_{ij}P_{ji} = \ln Q_{ij}Q_{ji} + C$ for all $(i,j)$. 
The effective forces (\ref{eff.force}) are thus invariant along an $e$ equivalence class,
\bea
X'_P = X'_Q \, .
\eea 

When looking at the entire $e$ class of $P$, it is convenient to parametrize its elements by the affinities $\pmb{A}$ instead of the parameters $\pmb{\lambda}$.
We then have that
\bea
Q^e (\pmb{A}) = s \left[ P^e \circ Z(\pmb{A}) \right]  \, .
\label{Qe.A}
\eea
Since the mapping $s$ is non-linear, the $e$ manifold depends non-linearly on the microscopic parameters (Figure \ref{fig.manifold}). 
It expands the concept of $e$-geodesic introduced in information geometry \cite{A16, FN07}.

\subsection{Transport and fluctuation symmetries of the $e$ scheme}

As discussed previously, the nonlinear response is meaningful only when both the macroscopic affinities and the complete set of microscopic parameters are specified along the path they trace in parameter space. 
Within the $e$ equivalence class, there exists a unique mapping between affinities (and currents) and the corresponding dynamics; all microscopic parameters are then determined by the appartenance to $[Q^e]$.
Transport is thus well-defined along an $e$ equivalence class.

We can thus examine the transport and fluctuation properties of the system under the $e$-control scheme (\ref{e.sol}).
We start by looking at the mean currents. 
By construction, the $e$-class spans the entire range of affinities with a unique equilibrium dynamics $P^e$. 
Also, since the class $Q^e$ always contains a dynamics and its time-reversed one, the current-affinity relationship will be anti-symmetric with respect to the affinities, 
\bea
J_\alpha (-\pmb{A}) = - J_\alpha(\pmb{A}) \, .
\eea
This relationship already imposes a symmetry that does not necessarily hold when varying a system's parameters along arbitrary paths \cite{A23}.

Next, we consider the current reponses (\ref{cur.resp}). As demonstrated in \cite{A22}, they satisfy the relation
\bea
\frac{\partial J_\alpha}{\partial A_{\beta} } (\pmb{A}) = \frac{\partial J_\beta}{\partial A_{\alpha} } (\pmb{A})\, ,
\label{NLR1}
\eea
demonstrating symmetry in indices ($\alpha, \beta$) for any $\pmb{A}$, i.e. both near and far from equilibrium \cite{A22}.
As a result, the response coefficients (\ref{J.exp}) are also symmetric in indices ($\alpha, \beta_1, ..., \beta_l$).
Moreover, they can be expressed in terms of equilibrium cumulants \cite{A22}:
\bea
L_{\alpha, \beta_1 ... \beta_l} &=& 0  \quad  \quad \quad \quad \quad \quad \quad  \quad \text{if {\it l} is even} \label{L.a} \\
L_{\alpha, \beta_1 ... \beta_l} &=& \parent{\frac{1}{2}}^l K_{\alpha \beta_1 ... \beta_l} (\pmb{0}) \quad \text{if {\it l} is odd.}
\label{L.b}
\eea 
In particular, this leads to the Onsager symmetry $L_{\alpha, \beta} = L_{\beta, \alpha}$ and the corresponding Green-Kubo formula $L_{\alpha, \beta} = (1/2) K_{\alpha\beta} (\pmb{0}).$ 
Similar expressions and symmetries can be derived for the higher-order cumulants and their nonequilibrium response.
Furthermore, all cumulants can be obtained by deriving the current-affinity manifold $J_\alpha (\pmb{A})$ \cite{A22}. 
Notablym the symmetries (\ref{L.a})-(\ref{L.b}) extend beyond those of the fluctuation theorem \cite{AG04, BG18, A22}. \\

Where does the symmetry (\ref{NLR1}) come from?
Its origin can be traced back to the expression (\ref{e.sol}) and realizing that $P\circ Z(\eta)$ forms the generator of the current generating function (\ref{CGF}).
Combining this observation with the representation (\ref{Qe.A}), the current generating function of $P$ can be expressed in terms of the equilibrium generating function of the {\it equilibrium} chain $P^e$:
\bea
G_{P} (\pmb{\eta}) = G_{P^e} (\pmb{\eta}+\pmb{A}/2) - G_{P^e} (\pmb{A}/2) \, .
\label{NEQfluct}
\eea
In other words, the nonequilibrium current fluctuations of $P$ are entirely determined by the equilibrium fluctuations of $P^e$ \cite{A12c}.
Since the equilibrium fluctuations are symmetric, these symmetries are shared by all elements of the corresponding $e$ class.

\subsection{Potential function of the $e$-class}

The symmetries (\ref{NLR1}) impose strong restrictions on the functions $J_\alpha (\pmb{A})$.
In particular, if we consider the vector field
\bea
\pmb{J} (\pmb{A}) = (J_{\alpha_1}, \cdots, J_{\alpha_M} ) \, ,
\eea
where $M = E+N-1$ is the number of independent currents, the conditions (\ref{NLR1}) imply the existence of a scalar function $\rho (\pmb{A}) $ - called the potential function - such that \cite{FN99}
\bea
\frac{\partial \rho (\pmb{A})}{\partial A_\alpha}  = J_\alpha (\pmb{A}) 
\eea
and where the kinetic forces $X_{ij}$ are kept constant.

For the $e$ control strategy, the potential function $\rho$ is given by  
\bea
\rho (\pmb{A}) = D (P^e|P(\pmb{A}) )
\label{DPPE}
\eea
between the equilibrium dynamics $P^e$ and $P = s[P^e \circ Z(\pmb{A})]$ \cite{A12c} and where
\bea
D(P|Q) = \sum_{ij} \pi_i P_{ij} \ln \frac{P_{ij}}{Q_{ij}} \geq 0 \, .
\label{KL}
\eea
is the KL divergence. The KL divergence is non-negative and vanishes only when $P$ and $Q$ are identical.
 
This formula further connects transport properties with information geometry concepts. 
In particular, the generating function (\ref{CGF}) is given the divergence (\ref{DPPE}) over the $e$ manifold.
This will be illustrated on an analytical example in Section \ref{MM}.

\subsection{Minimizing distortion in the control of mesocopic transport}

When controlling a system, it is often desirable to modify the device's outputs while remaining as close as possible to the original dynamics.
Remarkably, in addition to generating symmetric transport responses, the $e$ class also minimizes the distortion with respect to the original process.

Using the KL divergence as our measure of dissimilarity \cite{FN03}, we then have the following result:\\

\noindent {\bf Theorem}. The $e$ control scheme (\ref{e.sol}) solves the minimization problem
\bea
Q^e (\pmb{\lambda}) = 
\arg \min_Q \Big[ D(Q|P) + \sum_\alpha \lambda_\alpha J_\alpha[Q] \Big] \, ,
\label{FKL.min}
\eea
where the minimization is performed over the space of compatible chains \cite{FN02}.
\\

\noindent {\it Demonstration}: To obtain the solution (\ref{e.sol}), an explicit calculation can be performed using techniques from rate distortion theory, in particular the Blahut-Arimoto algorithm \cite{R72, S72, C06}. 
However, a simpler and more direct demonstration can be found in Ref. \cite{A12} based on a result from Ref. \cite{S11}. $\square$\\

In other words, the equivalence class (\ref{e.sol}) describes how to change the system's parameters to achieve a given output while minimizing the distortion $D(Q|P)$ with respect to the original system. 
This ensures that the adjustments made to the system are as targeted as possible, and minimize unrelated deviations or disruptions.

\begin{figure}[t]
\includegraphics[scale=.5]{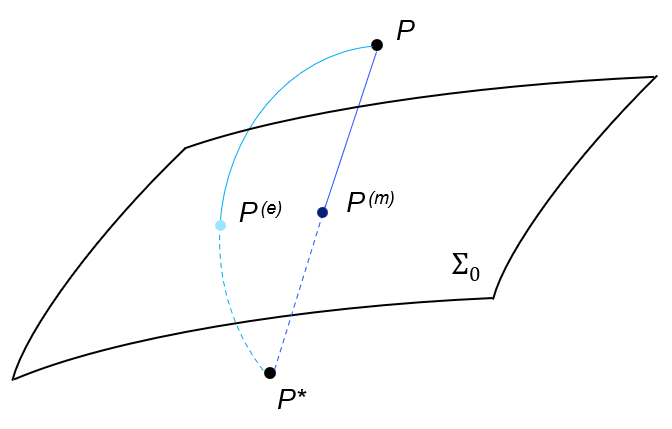}
\caption{ {\bf Schematic representation of the $e$ and $m$ equivalence classes in parameter space.}. 
Both $e$ and $m$ manifolds have dimension $E-N+1$ (here shown as 1-dimensional). 
They contain the time-reversed dynamics $P^*$, and cross the equilibrium manifold $\Sigma_0$ at $P^e$ and $P^m$, respectively. 
The equilibrium dynamics $P^e$ and $P^m$ are in general distinct from each other. 
The $e$-equivalence class typically forms a non-linear manifold while the $m$ class is linear in parameter space.
}
\label{fig.manifold}
\end{figure}


\section{m-control scheme}

\subsection{Current-based representation of Markov chains}

This representation generates Markov chains with given currents and activities, which will be used to define the $m$ equivalence classes. 
In this representation, we consider the flux matrix 
\bea
F_{ij} = \pi_i P_{ij} \, ,
\eea 
which describes the steady state probability fluxes between states.
$F$ satisfies
\bea
\sum_j F_{ij} = \pi_i \quad {\rm and} \quad \sum_i F_{ij} = \pi_j
\label{F.ss}
\eea
so that $\sum_{ij} F_{ij} = 1$.
Every such matrix $F$ corresponds to a unique chain $P$ and vice versa.

The currents (\ref{Jij}) and activities (\ref{Yij}) are obtained from the symmetric and anti-symmetric parts of $F$, respectively:
\bea
Y_{ij} &=& F_{ij} + F_{ij} \\
J_{ij} &=& F_{ij} - F_{ij} \, . 
\eea
The affinities of a cycle $c = (i_1, i_2, \cdots, i_n)$ can be calculated from the flux matrix as
\bea
A_c = \ln \parent{ \frac{F_{i_1i_2}\cdots F_{i_ni_1}}{F_{i_1i_2}\cdots F_{i_ni_1}} }
\label{Aff.rep}
\eea
since the stationary probabilities in the numerator and denominator cancel each other for a cyclic path.

The flux matrix can be decomposed as
\bea
F = \frac{(F+F^T)}{2} + \frac{(F-F^T)}{2} = \frac{Y+J}{2} \, .
\label{F.YJ}
\eea
In particular, the matrix $F$ is symmetric at equilibrium where $J=0$. 
More generally, the time-reversed dynamics $F^*$ is obtained by taking the transpose $F^T = (Y-J)/2$.
A flux matrix $F$ can also be decomposed as a convex combination of cycle matrices \cite{C81, A83, K94}.

From a flux matrix $F$ we can define the equilibrium system
\bea
\bar{F}
= \frac{F+F^T}{2} = \frac{Y}{2}\, .
\eea
The corresponding transition matrix
\bea
P^m = \frac{P+P^*}{2} \, ,
\label{Pm}
\eea
where we used that $\pi$ is the stationary distribution of both $P$ and $P^*$.
The chain $P^m$ is the equilibrium chain that has the closest projection distance to $P$ in the following information-theoric sense \cite{WW21}:
\bea
P^m = \arg \min_{Q \in \Sigma_0} D(P|Q) \, ,
\label{Pm.inf}
\eea
where the minimization is done over the space $\Sigma_0$ of compatible equilibrium chains. 
This anticipates the important role that $P^m$ will play in the characterization of the $m$ equivalence classes.
The chain $P^m$ is also related to the entropy production of the system \cite{A24b}.

\subsection{Definition of the $m$-control scheme}

We are now in position to define our $m$ control scheme.
In this scheme, the output of the system is controled by varying the currents $J_\alpha$.
Starting from the flux matrix $F_P$ associated with $P$, the {\bf $m$-control scheme} is defined as
\bea
F^m (\Delta \pmb{J}) = F_{P} + \frac{1}{2} \sum_\alpha  \Delta J_\alpha \ell_\alpha (C_\alpha - C_\alpha^T) \, .
\label{m.sol}
\eea
Here $\Delta J_\alpha = J'_\alpha - J_\alpha$ is the change in output current and $C_\alpha$ is the cycle matrix associated with the cycle $\alpha$ (Appendix A). 
The expression (\ref{m.sol}) is a flux matrix only when $0 \leq F^m_{ij} \leq 1$. 
The $m$ class is then obtained by varying the change in currents $\Delta \pmb{J}$ in their allowed range.

When looking at the entire $m$ class of $P$, it is convenient to parametrize its elements by the absolute currents instead of their variations $\Delta \pmb{J}$.
We then have that
\bea
F^m (\pmb{J}) = \bar{F} + \frac{1}{2} \sum_\alpha  J_\alpha \ell_\alpha (C_\alpha - C_\alpha^T) = \frac{Y}{2} + \frac{1}{2} \sum_\alpha  J_\alpha \ell_\alpha (C_\alpha - C_\alpha^T)\, .
\label{Qm.J}
\eea
The class $F^m$ forms a manifold with dimension equals to the number of independent currents.
From this expression, we see that the $m$ manifold contains the equilibrium chain $P^m$ and the time-reversed chain $P^*$.

The chains $F^m$ share similar characteristics for the symmetric part of their dynamics.
Indeed, using that the matrices $(C_\alpha-C_\alpha^T)$ are anti-symmetric and satisfy $\sum_j (C_\alpha-C_\alpha^T)_{ij} = 0$, we see that the stationary distribution $\pi$ is invariant as the $\pmb{J}$s are varied, and that $F + F^T$ is independent of $\pmb{J}$. 
That is, two dynamics $P$ and $Q$ in the same $m$ class have the same activities, 
\bea
Y_P = Y_Q \, ,
\label{m.def.alt}
\eea
and the same stationary distribution, $\sum_i \pi_i Q_{ij} = \pi_j$.

Since the stationary distribution $\pi$ is invariant along the equivalence class $m$, the class (\ref{Qm.J}) leads to the following transition probabilities:
\bea
Q^m_{ij} (\pmb{J}) = P^m_{ij} + \frac{1}{2\pi_i} \sum_\alpha  J_\alpha \ell_\alpha (C_\alpha - C_\alpha^T)\, .
\label{Pmij}
\eea
The parameters (transition probabilities) vary linearly with the currents. 
The $m$ control scheme is thus achieved by a linear variation of the parameters (Figure~\ref{fig.manifold}). 
The $m$ manifold expands the concept of $m$-geodesic introduced in information geometry \cite{A16, FN07}.

\subsection{Transport symmetries of the $m$-class}

Similar to the $e$ class, the $m$ class has well-defined transport properties as all the microscopic parameters are uniquely determined within the class. 
We can thus examine the transport properties of the system under the $m$-control scheme.

From expression (\ref{Aff.rep}), the affinity-current relationship for an $m$ class can be written as
\bea
A_\alpha (\pmb{J}) = \ln \prod_{e} \parent{ \frac{Y_e + \epsilon_\alpha (e) \sum_\rho J_\rho \epsilon_\rho (e) }{Y_e - \epsilon_\alpha (e) \sum_\rho J_\rho \epsilon_\rho (e)  } }  \, .
\label{AJ.exp}
\eea
Here, $\epsilon_\alpha (e) = \pm 1$ if $c_\alpha$ contains the edge $e$ in the positive ($+$) or negative orientation ($-$), and $0$ otherwise (Appendix A). In this expression, the activities $Y_e$ are fixed and the currents can vary in their allowed range.
These affinity-current relationships are anti-symmetric, i.e.
\bea
A_\alpha (-\pmb{J}) = - A_\alpha (\pmb{J}) \, .
\label{AJ.sym}
\eea
This symmetry doesn't necessarily hold when tracing an arbitrary path in parameter space \cite{A23}, indicating that the $m$ classes also display special transport properties.

Next, a direct calculation shows that the response takes the form
\bea
\frac{\partial A_\alpha}{\partial J_\beta } (\pmb{J}) = 2 \sum_e \frac{Y_e \, \epsilon_\alpha(e)\epsilon_\beta(e)}{Y_e^2 - (\sum_\rho \epsilon_\rho (e) J_\rho)^2} \, ,
\label{AJ.exp}
\eea
demonstrating symmetry in indices ($\alpha, \beta$):
\bea
\frac{\partial A_\alpha}{\partial J_\beta } (\pmb{J}) = \frac{\partial A_\beta}{\partial J_\alpha } (\pmb{J}) \, .
\label{AR.sym}
\eea
This symmetry holds for any $\pmb{J}$, i.e. both near and far from equilibrium.

In terms of the reponse coefficients (\ref{A.exp}), the symmetry (\ref{AR.sym}) imposes that all even-order coefficients are symmetric in the indices $(\alpha, \beta_1, \ldots, \beta_l)$ while all odd-order coeficients vanish.
Using the result (\ref{AJ.exp}), we can obtain exact expressions for the response coefficients: 
\bea
M_{\alpha, \beta_1 ... \beta_l} &=& 2 \sum_e \, \parent{\frac{1}{Y_e}}^l \epsilon_\alpha(e) \epsilon_{\beta_1}(e) \ldots \epsilon_{\beta_l}(e)  \quad\quad \text{if {\it l} is odd,} \\
M_{\alpha, \beta_1 ... \beta_l} &=& 0 \quad \quad \quad \quad\quad \quad\quad \quad\quad\quad \quad\quad \quad\quad \quad\quad \text{if {\it l} is even}.
\label{M.b}
\eea
The response coefficients are symmetric in the indices $(\alpha, \beta_1, \ldots, \beta_l)$ and are expressed in terms of the equilibrium activities.

These formulas reveal that coupling between processes is governed by the activities along the edges where both these reciprocal processes are connected. In particular, the coupling - at all order - between two processes will be lowered if the activities along these edges is increased, and vice-versa. 
This provides a direct and practical way to control the transport of mesoscopic systems, and will be illustrated in Section \ref{ION}.

They also reveal that the direct and coupling coefficients are linked: Because $\sum_e Y_e = 1$, it is not possible to change the coupling between two processes without affecting the self response coefficients $M_{\alpha\alpha}$.
\\

Where does the symmetry (\ref{AR.sym}) come from?
The symmetry of the nonlinear response emerge from the linear structure (\ref{m.sol}) and the structure of formula (\ref{AJ.exp}). In contrast to the $e$ equivalence class, there is no obvious connection to the underlying equilibrium fluctuations.

\subsection{Potential function of the $m$-class}

If we consider the vector field
\bea
\pmb{A} (\pmb{J}) = (A_{\alpha_1}, \cdots, A_{\alpha_M} ) \, ,
\eea
where $M = E+N-1$ is the number of independent affinities, the conditions (\ref{AR.sym}) imply that there exists a potential function $\mu (\pmb{J}) $ such that
\bea
\frac{\partial \mu (\pmb{J})}{\partial J_\alpha}   = A_\alpha (\pmb{J}) 
\eea
where the activities $Y_{ij}$ are kept constant.

An analytic form of the potential function $\mu$ remains to be derived.

\subsection{Minimizing distortion in the control of mesocopic transport}

In contrast with the $e$ control scheme, the $m$ class does not appear to follow a minimization principle similar to (\ref{FKL.min}), despite the structure suggested by (\ref{Pm.inf}). An interpretation of the $m$ control scheme in terms of information-theoric quantities remains to be explored.\\

\setlength{\fboxsep}{10pt}
\noindent
\fbox{
    \parbox{0.95\textwidth}{
        \textbf{The best of both worlds: When the $e$ and $m$ manifolds coincide} (Box 1)  \\

A special case occurs when the $e$ and $m$ classes are identical. 
When this happens, the system can be controled with a linear change of parameters while minimizing the distortion with respect to the original dynamics, and with the full range of possible currents (and with exact expressions for the affinity-current relationships).\\

The $e$ and $m$ equivalence classes coincide when $P^e = P^m$. 
For a one-dimensional ring, this occurs when $P_{i, i+1} = p = 1 - P_{i+1,1}$, i.e. for an homogeneous ring. 
The case of an arbitrary graph remains to be investigated.
    }
}


\section{Applications}

We now illustrate the construction and transport symmetries of the $e$ and $m$ equivalence classes on two systems.
We start by considering a molecular motor model with one independent current and a 2-dimensional parameter space. 
This allows us to provide analytical expressions for both the $e$- and $m$-classes and visualize their shapes in parameter space, as well as deriving the current-affinity curves.

Next, we consider a model of ions transport through a membrane with two coupled currents and a 8-dimensional parameter space.
While some analytical results are available, we focus on showing numerically that the symmetries (\ref{NLR1}) and (\ref{AR.sym}) hold far away from equilibrium.

\subsection{Molecular motor model} \label{MM}

We consider a Markov chain representing a molecular motor with $2\ell$ states corresponding to different conformations of the protein complex. 
These states form a cycle of periodicity $2\ell$ corresponding to a revolution by 360° for a rotary motor or a reinitialization step for a linear motor.
The motor alternates between two types of states according to the transition matrix \cite{AG06}
\bea
P=
\begin{pmatrix}
0 &  p_1 &  &  &  & 1-p_1\\
1-p_2 & 0 & p_2 &  &  & \\
 &  1-p_1 & 0 & p_1 & & \\
 &   & \ddots & \ddots & \ddots & \\
 &   &  & 1-p_1 & 0 & p_1\\
p_2 &   &  &  & 1-p_2 & 0
\end{pmatrix}_{2\ell \times 2\ell}  .
\nonumber
\eea
The matrix $P$ is doubly stochastic, so that its stationary state is given by the uniform distribution ${\bf \pi}= (1,1,\cdots ,1)/2\ell$ for all parameters $(p_1,p_2)$.
The independent current $J$ and affinity $A$ take the form
\bea
J = \frac{1}{2\ell} \parent{p_1+p_2 -1}
\label{MM.J}
\eea
and
\bea
A = \ell \, \ln \frac{p_1p_2}{(1-p_1)(1-p_2)} \, .
\label{MM.A}
\eea
The system is at equilibrium when $J=A=0$ (i.e., when $p_2 = 1-p_1$). 

The activities read $Y_1 = (p_1 -p_2 +1)/2\ell$ and $Y_2 = (p_2 -p_1 +1)/2\ell$ 
depending if the transition is odd or even, respectively. 
However, only one edge is independent due to the symmetry of the model and the constraint $Y_1+Y_2 =1/\ell$.
Taking 
\bea
Y = Y_1 = \frac{1}{2\ell}(p_1 -p_2 +1)
\label{MM.Y}
\eea
as the independent activity leads to an effective kinetic force
\bea
X = \ell \, (\bar{X}_1 - \bar{X}_2)  = \ell \, \ln \frac{p_1 (1-p_2)}{p_2 (1-p_1)} \, .
\label{MM.X}
\eea

\subsubsection{Transport along e-equivalence classes}

\begin{figure}[t]
\includegraphics[scale=.65]{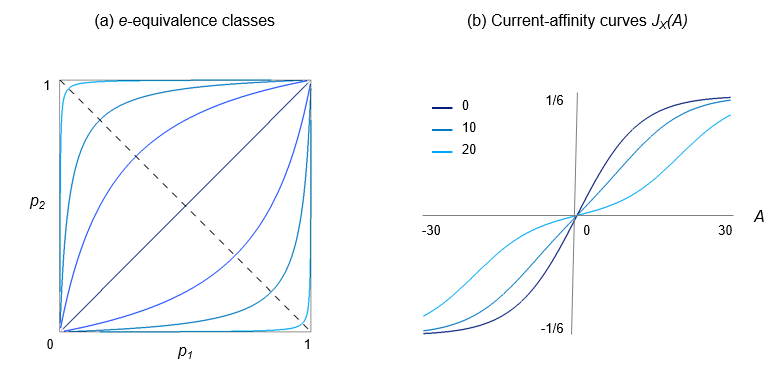}
\caption{ {\bf e-equivalence classes and corresponding current-affinity curves for the molecular motor}. 
(a) Equivalence classes for different values of $X =[-20, -10, -5, 0, 5, 10, 20]$. 
Each class traces a non-linear path in parameter space except for the class $X=0$.
Each class intersects the equilibrium manifold (dashed line) at a unique point $P_X^e$. 
(b) Current-affinity curves (\ref{MM.J.e}) for different equivalence classes $X$. 
Each curve is anti-symmetric and invariant under the transformation $X \rightarrow -X$.
Each curve spans the entire range of possible current values $\pm 1/2\ell$.
}
\label{fig.MM.e}
\end{figure}

The $e$ equivalence classes are obtained by varying the affinity $A$ at fixed value of $X$. 
Using expressions (\ref{MM.A}) and (\ref{MM.X}), the model parameters can be expressed as 
\bea
p_{1} = \frac{{\rm e}^{(A+X)/2\ell}}{1+{\rm e}^{(A+X)/2\ell}} \, , \quad \quad p_{2} = \frac{{\rm e}^{(A-X)/2\ell}}{1+{\rm e}^{(A-X)/2\ell}} \, . 
\label{MM.param} 
\eea
Therefore, each $e$ equivalence class traces a nonlinear path in parameter space (Figure \ref{fig.MM.e}a).

Along these equivalence classes, the current-affinity curves take the form
\bea
J_X(A) = \frac{1}{2\ell} \frac{({\rm e}^{A/\ell} -1)}{(1+{\rm e}^{(A+X)/2\ell})(1+{\rm e}^{(A-X)/2\ell})} \, .
\label{MM.J.e}
\eea
Within these equivalence classes, the current-affinity curves are well-defined (Figure \ref{fig.MM.e}b). 
The currents are anti-symmetric, reach the maximal and minimal possible values $\pm 1/2\ell$ when the affinity reaches $\pm \infty$, and are invariant under the transformation $X \rightarrow -X$.
The current-affinity curves (\ref{MM.J.e}) also capture fluctuation at all orders, i.e. deriving $J_X(A)$ with respect to $A$ generates the various cumulants for any value of the affinity (not shown) \cite{A23}.

The potential function is given by \cite{A24b}
\bea
\rho = D(P^e|P) = - \ln \parent{\sqrt{p_1(1-p_2)} + \sqrt{p_2(1-p_1)}}
\label{MM.rho}
\eea
with $P^e$ given by
\bea
P^e = \frac{1}{\Sigma} (\sqrt{p_1 (1-p_2)} , \sqrt{p_2 (1-p_1)}) \, ,
\label{MM.Pe}
\eea
where $\Sigma = \sqrt{p_1 (1-p_2)} + \sqrt{p_2 (1-p_1)}$. 
Inserting expressions (\ref{MM.param}) into (\ref{MM.rho}), the current $J$ in the $e$ manifold is obtained by deriving with respect to $A$ (and keeping $X$ constant). 
The potential function $\rho(A)$ is also the generating function of the currents of dynamics $P$.

The $e$ manifold also minimize the divergence (\ref{KL}), thus allowing to tune the output of the system while minimizing the other dynamical variations.


\subsubsection{Transport along m-equivalence classes}

\begin{figure}[t]
\includegraphics[scale=.65]{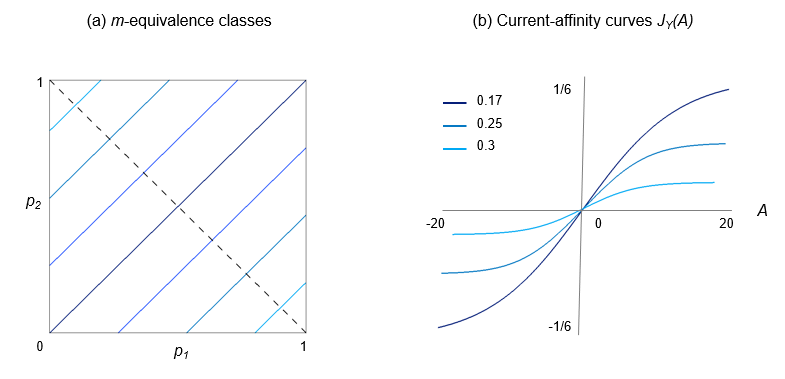}
\caption{ {\bf $m$ equivalence classes and corresponding current-affinity curves for the molecular motor}.
(a) $m$ equivalence classes for different values of $Y=Y_1 = [0.03, 0.08, 0.12, 0.17, 0.21, 0.26, 0.3]$. 
Each class forms a straight line in parameter space, and intersects the equilibrium manifold (dashed line) at a unique point $P_Y^m$. 
(b) Current-affinity curves (\ref{MM.A.m}) for different equivalence classes $Y$. 
Each curve is anti-symmetric and invariant under the transformation $Y \rightarrow 1/\ell-Y$.
The currents reach different limiting values depending on the value of $Y$.
}
\label{fig.MM.m}
\end{figure}

The $m$-equivalence classes can be obtained from the construction (\ref{Qm.J}). 
Since the system has a uniform steady state, the flux matrix $F = P/2\ell$.
It can be decomposed as
\bea
F= \frac{1}{2\ell}
\begin{pmatrix}
0 &  p_1 &  &  &  & 1-p_1\\
1-p_2 & 0 & p_2 &  &  & \\
 &  1-p_1 & 0 & p_1 & & \\
 &   & \ddots & \ddots & \ddots & \\
 &   &  & 1-p_1 & 0 & p_1\\
p_2 &   &  &  & 1-p_2 & 0
\end{pmatrix} 
=
\frac{1}{2} \begin{pmatrix}
0 &  Y_1 &  &  &  & Y_2\\
Y_1 &  & Y_2 &  &  & \\
 &  Y_2 &  & Y_1 & & \\
 &   & \ddots & 0 & \ddots & \\
 &   &  & Y_2 & 0 & Y_1\\
Y_2 &   &  &  & Y_1 & 0
\end{pmatrix}  
+ \frac{J}{2}
\begin{pmatrix}
0 &  1 &  &  &  & -1\\
-1 &  & 1 &  &  & \\
 &  -1 &  & 1 & & \\
 &   & \ddots & 0 & \ddots & \\
 &   &  & -1 & 0 & 1\\
1 &   &  &  & -1 & 0
\end{pmatrix} 
\nonumber
\eea
The $m$ equivalence classes are obtained by fixing the activities $Y_i$ (remember that $Y_1 +Y_2 = 1/\ell$) and varying the current $J$.
Since the transition probabilities satisfy $0 \leq p_i \leq 1$, the current varies between $-\min(Y_1, Y_2) \leq J \leq \min(Y_1, Y_2)$.
In terms of the $(p_1, p_2)$ variables, we have
\bea
(p_1, p_2) = \ell \, (Y_1 + J, Y_2 + J) \, .
\eea
That is, the $m$ equivalence classes form straight lines in the parameter space (Figure \ref{fig.MM.m}a) \cite{FN04}.
The $m$ equivalence class for the spatially homogeneous system $p_1 = p_2$ or $Y_1 = Y_2 = 1/2\ell$ coincides with the corresponding $e$ equivalence class (Figures \ref{fig.MM.e} and \ref{fig.MM.m}, see also Box 1). 

The affinity-current curves are well-defined in the equivalence classes and take the form
\bea
A_Y(J) = \ell \ln \left[\parent{ \frac{Y_1 + J}{Y_1 - J}} \parent{\frac{Y_2 + J}{Y_2 - J} }\right]
\label{MM.A.m}
\eea
where the $Y_i$ are fixed. 
The response curves are monotonic, anti-symmetric in $J$, and invariant under the transformation $Y \rightarrow 1/\ell-Y$. 
However, in contrast with the $e$-classes, the currents do not span the entire possible range $\pm 1/2\ell$ (Figure \ref{fig.MM.m}b).

Using formulas (\ref{AJ.exp}), we see that the response $dA/dJ$ is largest when $Y_1 = Y_2$, for all value of $J$ (Figure \ref{fig.MM.m}b). In this control scheme, transport of one-dimensional systems can thus be controlled by modifying the activities.

While not a general property of $m$ classes, in this system they satisfy a minimization principle akin to (\ref{FKL.min}) but for the reversed divergence $D(P|Q)$ instead of $D(Q|P)$ \cite{M12}.


\subsection{Model of ions transport across a membrane} \label{ION}

\begin{figure}[h]
\centerline{\includegraphics[width=12cm]{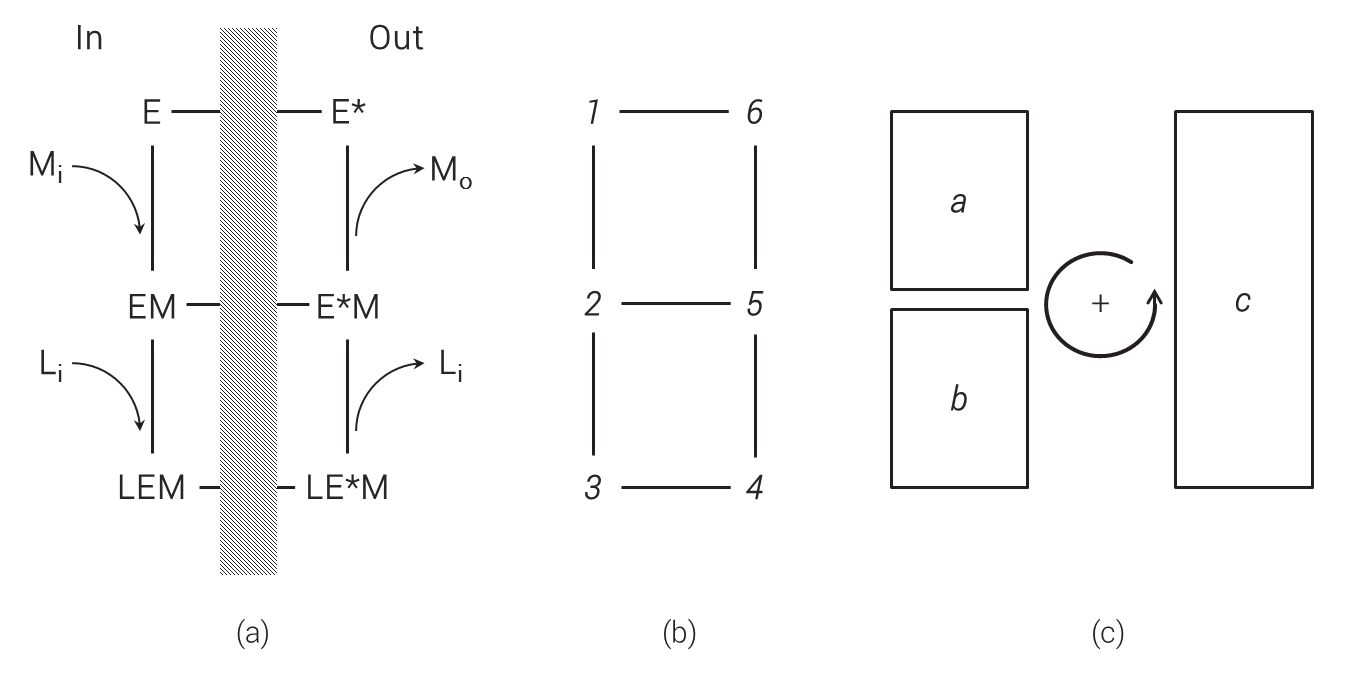}}
\caption{{\bf Ion transport model with two independent currents}. 
(a) Mechanism for the transport of ions M and L across the membrane. 
In the normal mode of operation, molecule M has a larger concentration inside than outside, $[{\rm M_i}] > [{\rm M_o}]$, while the opposite holds for molecule L, $[{\rm L_o}] > [{\rm L_i}]$. 
The complex E acts as a free energy transducer and utilizes the M concentration gradient to drive molecules of L from inside to outside against its concentration gradient. 
For example, in the case of the Na/K-ATPase complex M and L would correspond to ${\rm K}^+$ and ${\rm Na}^+$, and transport would be coupled to ATP consumption. (b) Kinetic diagram. (c) Cycle decomposition. The cycles $a$ and $b$ are chosen as the two independent cycles. The positive orientation is chosen counterclockwise (adapted from Hill \cite{H05}).}
\label{fig.ION.diag}
\end{figure}

Let's now turn to a model of ion transport as an archetype of systems with two independent currents. 
Following Hill \cite{H05}, consider a cell surrounded by a membrane that separates the cell's interior (In) from its environment (Out). 
A complex E, which can exist in two conformations E and E*, has binding sites for ions L and M.
These sites are accessible to inside molecules in configuration E only, and to outside molecules in configuration E* only. 
L can be bound only when M is already bound on its site (Figure \ref{fig.ION.diag}a).

The system has $N=6$ states connected by the kinetic diagram depicted in Figure~\ref{fig.ION.diag}b.
The parameter space $\Lambda$ is $8$-dimensional while the set of equilibrium dynamics forms an $6$-dimensional manifold. 
There are two independent affinities or currents, here measured along cycles $a$ and $b$ (Figure~\ref{fig.ION.diag}c). 

\subsubsection{Transport along the e-equivalence classes}

\begin{figure}[h]
\centerline{\includegraphics[width=10cm]{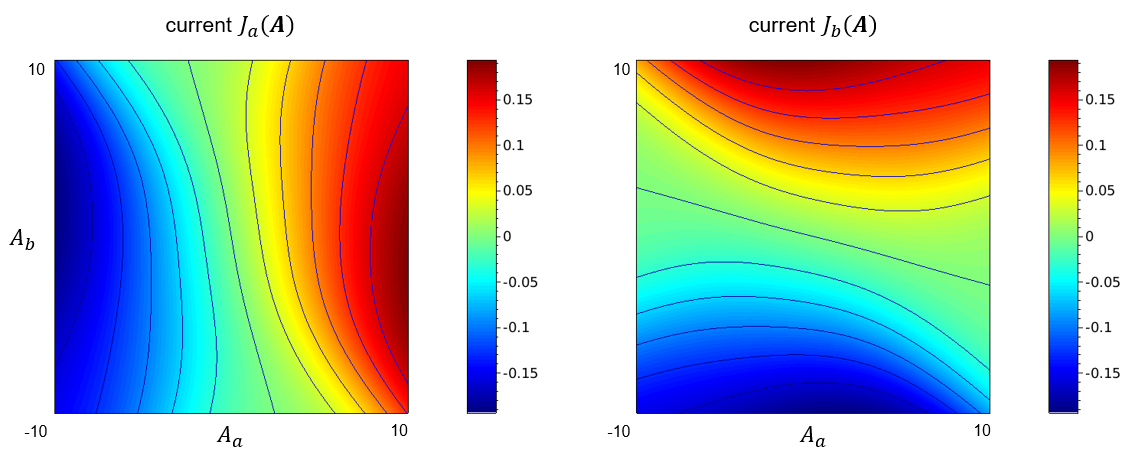}}
\caption{{\bf Coupled currents of the ion transport model along an $e$ equivalence class.} 
Currents $J_a$ and $J_b$ and their isolines (solid curves) as a function of the affinities $A_a$ and $A_b$ in a given equivalence class. The equivalence class is defined by the equilibrium dynamics $P^e$ with $p_{12} = 0.62, p_{16} = 0.38,p_{21} = 0.57, p_{23} = 0.13, p_{25} = 0.30, p_{32} = 0.41, p_{34} = 0.59, p_{43} =0.46, p_{45} = 0.54, p_{52} = 0.42, p_{54} = 0.32, p_{56} = 0.26, p_{61} = 0.66,$ and $p_{16} = 0.34$. 
}
\label{fig.ION.e}
\end{figure}

We illustrate the transport dynamics in a given equivalence class $e$. 
Each equivalence class corresponds to a $2$-dimensional manifold in the $8$-dimensional parameter space.
As expected, these currents are coupled and nonlinear, i.e. a change in $A_a$ will drive the current $J_b$ and vice versa (Figure~\ref{fig.ION.e}). 
Whithin an $e$ equivalence class, the anti-symmetry $J_\alpha(-\pmb{A}) = - J_\alpha(\pmb{A})$ of the current-affinity manifold is verified. 

\begin{figure}[h]
\centerline{\includegraphics[width=10.5cm]{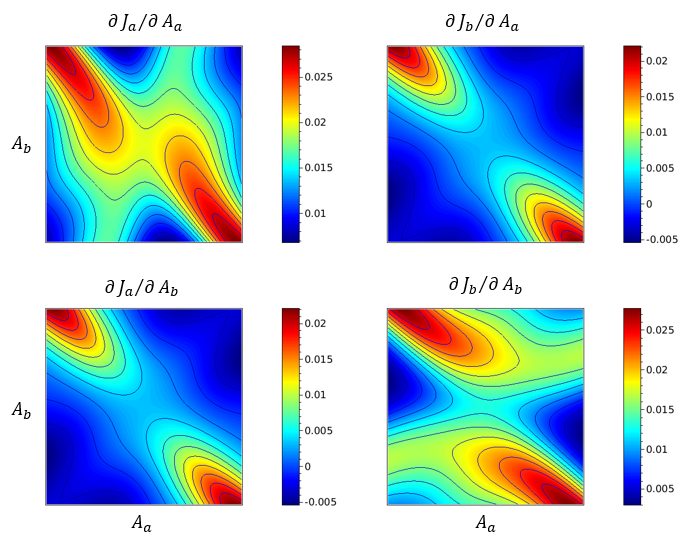}}
\caption{{\bf The currents response is symmetric both in the linear and nonlinear regimes.} The response coefficients $\partial \pmb{J}/\partial \pmb{A}$ display a complex structure in the nonlinear regime. 
Nonetheless, the cross-response coefficients $\partial J_a/\partial A_b = \partial J_b/\partial A_a$ are identical both near and far from equilibrium. 
Parameters take the same values as in Figure~(\ref{fig.ION.e}).}
\label{fig.ION.esym}
\end{figure}

In addition, and despite their nonlinear behavior, the currents responses display the 'hidden symmetry' (\ref{NLR1}) both near and far from equilibrium \cite{A24a}:
\bea
\frac{\partial J_a}{\partial A_b} (\pmb{A}) = \frac{\partial J_b}{\partial A_a} (\pmb{A})
\label{IT.sym}
\eea
{\it for all values of the affinities $\pmb{A}$} in a given equivalence class (Figure \ref{fig.ION.esym}). 
The symmetry (\ref{IT.sym}) is also confirmed for all other $e$ equivalence classes (not shown).
Higher-order fluctuations can be obtained by further derivation of the currents and their responses are also symmetric \cite{A22}. 


\subsubsection{Transport along the m-equivalence classes}

\begin{figure}[h]
\centerline{\includegraphics[width=10cm]{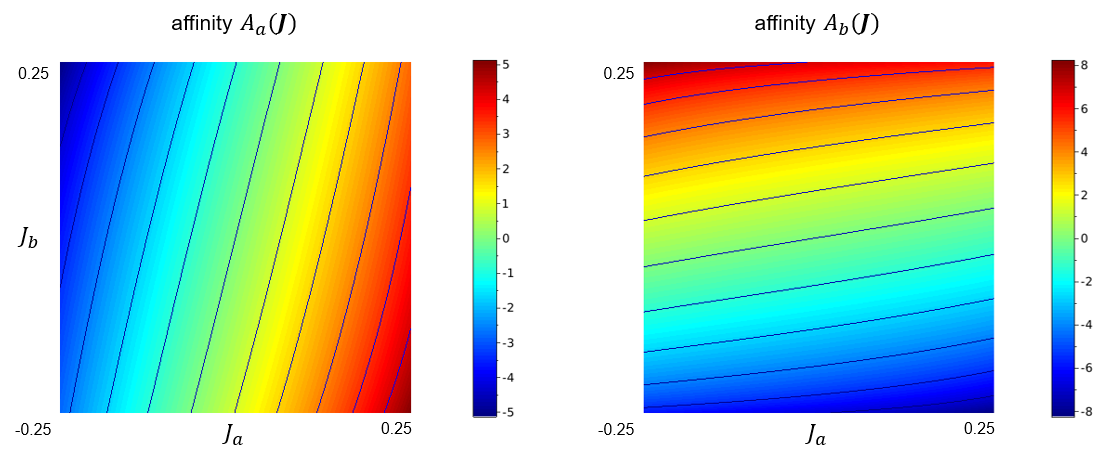}}
\caption{{\bf Affinities of the ion transport model along an $m$ equivalence class.} 
Affinities $A_a$ and $A_b$ and their isolines (solid curves) as a function of the currents $J_a$ and $J_b$ in a given equivalence class. 
The equivalence class is defined by the equilibrium dynamics $P^m$ with the same parameters as in Figure~\ref{fig.ION.e}.}
\label{fig.ION.m}
\end{figure}

Each $m$ equivalence class also corresponds to a $2$-dimensional manifold in the $8$-dimensional space.
The $m$ classes also display a nonlinear behavior and, beyond the anti-symmetry $A_\alpha(-\pmb{J}) = - A_\alpha(\pmb{J})$ characteristic of equivalence classes, no other symmetry is apparent (Figure \ref{fig.ION.m}).

\begin{figure}[h]
\centerline{\includegraphics[width=10.5cm]{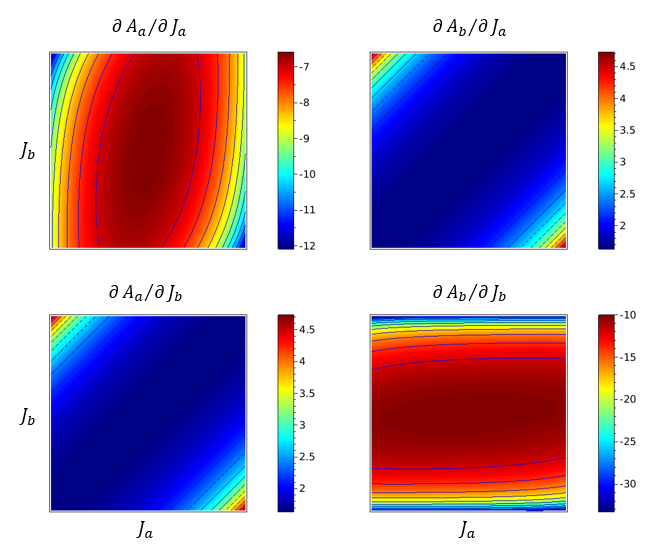}}
\caption{{\bf The affinity response is symmetric both in the linear and nonlinear regimes.} The response coefficients $\partial \pmb{A}/\partial \pmb{J}$ display a complex structure in the nonlinear regime. 
Nonetheless, the cross-response coefficients $\partial A_a/\partial J_b = \partial A_b/\partial J_a$ are identical both near and far from equilibrium. 
Parameters take the same values as in Figure~(\ref{fig.ION.m}).}
\label{fig.ION.msym}
\end{figure}

Despite the nonlinearities, the affinity-current responses display the 'hidden symmetry' (\ref{AR.sym}) both near and far from equilibrium:
\bea
\frac{\partial A_a}{\partial J_b} (\pmb{J}) = \frac{\partial A_b}{\partial J_a} (\pmb{J})
\label{IT.sym}
\eea
{\it for all values of the currents $\pmb{J}$} in a given equivalence class (Figure~\ref{fig.ION.msym}). 
The symmetry (\ref{IT.sym}) is also confirmed for all other $m$ equivalence classes.
The parameters are varied lineraly to generate this equivalence class (not shown).

In this model, the two transport processes are coupled via the edge $2 \longleftrightarrow 5$. According to our result (\ref{M.b}), the activity $Y_{25}$ directly determines the strength of the coupling between the two processes; in particular two systems with the same activity $Y_{25}$ will have the same cross-response for all values of the currents (not shown).
This illustrates how our results can lead to better control and design of mesoscopic devices.


\setlength{\fboxsep}{10pt}
\noindent
\fbox{
    \parbox{0.95\textwidth}{
        \textbf{What about time-dependent driving?} (Box 2) \\
        
This study focuses on the properties of the $e$ and $m$ equivalence classes in the context of steady-state transport. 
An important question arises: Do these equivalence classes play a role under a time-dependent driving?\\

While a comprehensive investigation of this scenario is beyond the scope of this paper, it is noteworthy that the $m$ classes exhibit another significant feature under time-dependent conditions. 
Specifically, if the external driving keeps the parameters within a given $m$ class, the system will instantaneously adapt and generate the desired currents. 
This phenomenon occurs because the steady state remains invariant within an $m$ class.\\

Indeed, consider a system driven by a time-dependent force. The probability distribution evolves in time as $e(n+1) = e(n) P(n)$ where $P(n)$ is determined by the external driving. The instantaneous currents then read $J_{ij} (n) = e_i(n)P_{ij}(n)-e_j(n)P_{ji}(n)$. If $\pi(n)$ is the stationary distribution of the dynamics $P(n)$, the instantaneous currents can also be written $J_{ij}(n) = \bar{J}_{ij}(n) + \Delta J_{ij}(n)$, where $\bar{J}_{ij}(n)$ is the steady state current under dynamics $P(n)$, and where the term $\Delta J_{ij}(n) = [e_i(n)-\pi_i(n)]P_{ij}(n)-[e_j(n)-\pi_j(n)]P_{ji}(n)$ reflects the contribution from the non-stationary part of the probability distribution. 
However, within an $m$ class the stationary distribution is invariant, i.e. $\pi = \pi P(n)$ for all $P(n)$ in the $m$ class. Therefore, if the initial distribution is the invariant distribution $e(0) = \pi$, the term $\Delta J_{ij}(n) = 0$ at all times and the instantaneous currents read $J_{ij}(n) = \bar{J}_{ij}(n)$. 
That is, they are given by the steady state currents from the controled dynamics $P(n)$.
This property holds for continuous-time Markov processes, making it directly applicable to many systems.
    }
}



\section{Summary and outlook}

In this study, we present a novel framework for controlling mesoscopic transport. 
By systematically varying combinations of the system's parameters, we can effectively regulate its output while ensuring symmetric transport properties out of equilibrium.
Achieving symmetric transport properties away from equilibrium simplifies the behavior of the system and opens new avenues for apprehending, designing and controlling mesoscopic systems.

We employ two geometric structures as a natural way to navigate the parameter space \cite{A16}, leading to the development of two distinct control schemes.
Each scheme generates a corresponding dynamical equivalence class, denoted as $e$ and $m$, which describes how parameter adjustments can achieve a desired output.
In both schemes the system's response is symmetric for all reciprocal transport processes even far from equilibrium, as illustrated on two biophysical systems. 

These transport symmetries arise through different mechanisms in the $e$ and $m$ classes. 
While both influence the anti-symmetric components of the dynamics, the symmetric components are constrained differently in each case. 
Indeed, an $e$-class fixes the kinetic forces while an $m$-class maintains constant activities \cite{FN03}.
Future research could explore different schemes for generating symmetric transport properties, and investigate potential connections with other frameworks, as explored in \cite{V18, V19, MW06, AE24, A24c}. 
In addition, the role of these equivalence classes in non-stationary scenarios, such as systems relaxing towards a steady state or driven by time-dependent forces, remain to be investigated (but see Box 2 for first insights).
 
Despite their conceptual similarities, one scheme might be better suited than the other for a specific system or application, depending on one's focus and ability to control the parameters of the system.
One the one hand, the $e$ control scheme presents useful properties, including enhanced symmetries for higher-order fluctuations and the capacity to achieve the full range of possible currents.
It also offers the advantage of modifing the device's outputs while remaining as close as possible to the original dynamics, thereby minimizing unrelated deviations or disruptions.
However, the $e$ necessitates nonlinear parameter changes for system control. 

The $m$ control scheme, on the other hand, allows for linear parameter adjustments to tune the output, with affinity-current relationships that can be calculated analytically. 
However, the range of achievable currents is limited in each equivalence class and does not encompass the full spectrum of possible currents.

These equivalence classes serve as natural extensions of information geometric concepts in non-equilibrium contexts. 
In the linear regime, the parameter space exhibits a well-defined structure based on the Fisher information metric, which is associated with equilibrium fluctuations \cite{SC12, ZS12}. 
However, the structure of the parameter space outside the linear regime remains largely unexplored, and our classes provide new frameworks with clear thermodynamic interpretations \cite{A24b, QEA23}.
The impact of alternative dissimilarity measures \cite{C10, A16,A24c} on transport properties also deserves further examination.

In practical terms, nonlinear reciprocity relations can improve the design and optimization of devices that operate under strong nonequilibrium conditions. For instance, advanced thermoelectric materials, chemical reactors, or energy conversion systems could benefit from design principles that exploit these extended symmetries to maximize efficiency or stability. This is seen for example in mass separation by effusion, which satisfies the symmetries (\ref{NLR1}) and whose maximal efficiency is achieved in the nonlinear regime \cite{GA11}. 
Full symmetries have also been observed in chemical reactions \cite{AG08} and transistors \cite{GG20}.
In this direction, our analytical result (\ref{M.b}) shows how to tune the coupling between processes based on the activities of selected edges. 
Similarly, feedback systems designed for devices like high-performance batteries or catalytic reactors could use real-time data to maintain optimal performance by exploiting predictable coupled responses — even when the devices are pushed beyond their standard range of operations.

The nonlinear reciprocity relations also significantly constrain the form of nonlinear transport equations, requiring that they derive from a scalar potential function and vastly reducing the number of independent parameters. This could simplify the analysis and modeling efforts of complex systems such as turbulent flows, biological processes, or strongly driven materials.
 
Finally, the principle of adapting a system under a minimization constraint can be extended to meet various objectives. 
For instance, a framework for simplifying generative models of stochastic processes has been developed by adapting the entropy rate under a dissimilarity constraint \cite{HK16}. 
Potential applications include enhancing model quality for sampling tasks and achieving clean models from corrupted data (non-parametric denoising) \cite{FN01}. 
These concepts could unlock new applications in the modeling and characterization of mesoscopic systems.\\

{\bf Acknowledgments.} I thank the Solvay Institutes for the workshop organized in honor of Prof. Pierre Gaspard, where some of this research was initiated. 
I'm also grateful to David Lacoste for encouraging me to write down these results.
\\

\centerline{\rule{7cm}{0.4pt}}

\vskip 0.2 cm

\section*{Appendix A: Fundamental currents and affinities}

Schnakenberg’s theory decomposes the thermodynamical properties of stochastic dynamics into their independent
contributions \cite{S76}.

A graph $G$ is associated with a stochastic dynamics as follows: each state i of the system corresponds to a vertex
or node while the edges represent the different transitions $i \rightarrow j$ allowed between the states.
An orientation is given to each edge of the graph $G$. The directed edges are thus defined by $i \rightarrow j$.
A graph $G$ usually contains cyclic paths. However, not all such paths are independent. They can be expressed
by a linear combination of a smaller subset of cycles, called the fundamental set, which plays the role of a basis in
the space of cycles. To identify all the independent cycles of a graph we introduce a maximal tree $T(G)$, which is a
subgraph of $G$ that satisfies the following properties:
\begin{itemize}
    \item $T$ contains all the vertices of $G$;
    \item $T$ is connected;
    \item $T$ contains no circuit, i.e., no cyclic sequence of edges.
\end{itemize}
In general a given graph $G$ has several maximal trees.

The edges $l$ of $G$ that do not belong to $T$ are called the chords of $T$. 
For a graph with $N$ vertex and $E$ edges, there exists $M=E-N+1$ chords. 
If we add to $T$ one of its chords $l$, the resulting subgraph $T + l$ contains exactly one circuit, $c_l$, which is obtained from $T + l$ by removing all the edges that are not part of the circuit. 
Each chord $l$ thus defines a unique cycle $c_l$ called a {\it fundamental cycle}. 
In this paper, we use the convention that the orientation is such that the cycles are oriented as the chords $l$.

We can formulate many important thermodynamic concepts in terms of cycles. 
For instance, the affinity of an arbitrary cycle $c$ can be expressed as a linear combination of the affinities of a fundamental set:
\bea
A(c) = \sum_l \epsilon_l (c) A(c_l) \, .
\eea
where the sum extends over all the chords, and $\epsilon_l (c) = \pm 1$ if $c$ contains the edge $l$ in the positive ($+$) or negative orientation ($-$), and $0$ otherwise.
Accordingly, the maximal tree $T$
can be chosen arbitrarily because each cycle $c_l$ can be redefined by
linear combinations of the fundamental cycles.
The fundamental cycles thus constitute a basis identifying the {\it independent} contributions to the stochastic process.

If multiple cycles $c_l$ generate the same macroscopic currents or affinities, we then group them into a common to obtain the resulting macroscopic quantities, e.g. $J_\alpha = \sum_{l \in \alpha} J_l$.

We can also express these relationships in terms of matrices. 
If the cycle $\alpha$ is the sequence of distinct integers $[a_1, \ldots, a_m]$, then the corresponding cycle matrix $C_\alpha$ is the $N \times N$ matrix $C_[a_1,\ldots ,a_m]$ given by $C_{a_1a_2} = C_{a_2a_3} = \ldots = C_{a_ma_1} = 1/m$ and $0$ otherwise. The length of cycle $\alpha$ is $\ell_\alpha = m$. The entries of any such cycle matrix sum to one.

Using these cycle matrices, we have that
\bea
J_{ij} = \sum_\alpha J_\alpha \ell_\alpha (C_\alpha - C_\alpha^T)_{ij} = \sum_\alpha \gamma_{ij}(c_\alpha)J_\alpha \, ,
\eea
where we introduce the matrices $\gamma (c_\alpha) = \ell_\alpha (C_\alpha - C_\alpha^T)$.
The matrices $\gamma (c_\alpha)$ are anti-symmetric while Kirchoff's law (\ref{Kirch}) imposes 
\bea
\sum_j \gamma_{ij}(c_\alpha) = \ell_\alpha \sum_j (C_\alpha - C_\alpha^T)_{ij} = 0 \, ,
\eea
and similarly when summing over the indices $i$.


\begin{thebibliography}{99}

\bibitem{H05} T. L. Hill, {\it Free Energy Transduction and Biochemical Cycle Kinetics} (Dover, 2005).

\bibitem{GEA07} R. Gutenkun, J. Waterfall, F. Casey, et al., {\it Universally sloppy parameter sensitivities in systems biology models}, PLoS Comput Biol {\bf 3}: e189 (2007).

\bibitem{MEA13} B. B. Machta, R. Chachra, M. K. Transtrum and J. P. Sethna, {\it Parameter Space Compression Underlies Emergent Theories and Predictive Models}, Science {\bf 342}, 604 (2013).

\bibitem{O31} L. Onsager, {\it Reciprocal Relations in Irreversible Processes. I.}, Phys. Rev. {\bf 37}, 405 (1931).

\bibitem{GM11} S. R. de Groot and P. Mazur, {\it Non-Equilibrium Thermodynamics} (Dover, 2011).

\bibitem{NP77} G. Nicolis and I. Prigogine, {\it Self-Organization in Nonequilibrium Systems} (Wiley, 1977).

\bibitem{V18}  H. Vroylandt, D. Lacoste and G. Verley, {\it Degree of coupling and efficiency of energy converters far-from-equilibrium}, J. Stat. Mech. 023205 (2018).

\bibitem{P67} R. Peterson, {\it Formal theory of nonlinear response}, Rev. Mod. Phys. {\bf 39}, 69 (1967).

\bibitem{AG04} D. Andrieux and P. Gaspard, {\it Fluctuation theorem and Onsager reciprocity relations}, J. Chem. Phys. {\bf 48}, 571 (2004).

\bibitem{BG18} M. Barbier and P. Gaspard, {\it Microreversibility, nonequilibrium current fluctuations, and response theory}, J. Phys. A {\bf 51}, 355001 (2018).

\bibitem{A12c} D. Andrieux, {\it Nonequilibrium large deviations are determined by equilibrium dynamics}, arXiv:1212.1807 (2012).

\bibitem{A24c} D. Andrieux, {\it A Minkowski space embedding to understand Markov models dynamics}, Phys. Rev. E {\bf 111}, 054109 (2025).

\bibitem{S76} J. Schnakenberg, {\it Network theory of microscopic and macroscopic behavior of master equation systems}, Rev. Mod. Phys {\bf 48}, 571 (1976).

\bibitem{MW06} C. Maes and M. H. van Wieren, {\it Time-symmetric fluctuations in nonequilibrium systems}, Phys. Rev. Lett. {\bf 96}, 240601 (2006).

\bibitem{TQS20} H. K. Teoh, K. N. Quinn, J. Kent-Dobias, et al., {\it Visualizing probabilistic models in Minkowski space with intensive symmetrized Kullback-Leibler embedding}, Phys. Rev. Res. {\bf 2}, 033221 (2020).

\bibitem{A23} D. Andrieux, {\it Making sense of nonequilibrium current fluctuations: A molecular motor example}, arxiv:2306.01445 (2023).

\bibitem{T64} K. Tani, {\it A formula of non-linear responses}, Prog. Theor. Phys {\bf 32}, 167 (1964).

\bibitem{V19} H. Vroylandt, D. Lacoste and G. Verley, {\it An ordered set of power-efficiency trade-offs}, J. Stat. Mech. 054002 (2019).

\bibitem{FN01} The representation (\ref{e.rep}) allows a direct parametrization in terms of the affinities.
More generally, we can generate a Markov chain with given affinities and kinetic forces. Introducing the symmetric matrix $W_{ij} (\pmb{X}) = \exp(X_{ij}/2)$, the chain $s[W(\pmb{X})\circ Z(\pmb{A})]$ has affinities $\pmb{A}$ and kinetic forces $\pmb{X}+C$. Since the kinetic forces are defined up to a constant, we can generate the entire space of Markov chains by varying these parameters.
However, while the affinities are directly available in this representation, no simple expression exists for the corresponding currents.

\bibitem{A16} S. Amari, {\it Information Geometry and Its Applications} (Springer, 2016).

\bibitem{M12} For an clear discussion on the difference between the divergences $D(P|Q)$ and $D(Q|P)$, see K. P. Murphy, {\it Machine Learning: A Probabilistic Perspective} (MIT Press, 2012).

\bibitem{A12} D. Andrieux, {\it Equivalence classes for large deviations}, arXiv:1208.5699 (2012).

\bibitem{N05} H. Nagaoka, {\it The exponential family of Markov chains and its information geometry}, The proceedings of the Symposium on Information Theory and Its Applications {\bf 28}, 601 (2005). 

\bibitem{WW21} G. Wolfer and S. Watanabe, {\it Information Geometry of Reversible Markov Chains}, Information Geometry {\bf 4}, 393 (2021).

\bibitem{FN06} Using a similar approach, we can generate a Markov chain with given affinities and kinetic forces. Introducing the symmetric matrix $W_{ij} (\pmb{X}) = \exp(X_{ij}/2)$, the chain $s[W(\pmb{X})\circ Z(\pmb{A})]$ has affinities $\pmb{A}$ and kinetic forces $\pmb{X}+C$. Since the kinetic forces are defined up to a constant, this completes the construction.

\bibitem{C81} J. E. Cohen, {\it A geometric representation of stochastic matrix: Theorem and conjecture}, The Annals of Probability {\bf 9}, 899 (1981).

\bibitem{A83} S. Alpern, {\it Rotational representations of stochastic matrices}, The Annals of Probability {\bf 11}, 789 (1983).

\bibitem{K94} S. Kalpazidou, {\it Rotational representations of transition matrix functions}, The Annals of Probability {\bf 22}, 703 (1994).


\bibitem{FN02} A stochastic matrix $Q$ is said to be compatible with $P$ if it satisfies $Q_{ij} \geq 0$ if $P_{ij} \geq 0$ and $Q_{ij} = 0$ if $P_{ij} = 0$.

\bibitem{A24b} D. Andrieux, {\it Irreversibility as divergence from equilibrium}, J. Stat. Phys {\bf 192}, 146 (2025).

\bibitem{V22} G. Verley, {\it Dynamical equivalence classes for Markov jump processes}, J. Stat. Mech. 023211 (2022).

\bibitem{FN07} The $e$ geodesic between $P$ and $P^e$ is obtained by varying all parameters $\lambda_\alpha$ uniformly, $\lambda_\alpha = \lambda$ for all $\alpha$. Similarly, the $m$ geodesic between $P$ and $P^m$ is obtained by varying all currents $J_\alpha$ uniformly, $J_\alpha = J$ for all $\alpha$.

\bibitem{FN03} If $P$ represents the true underlying process and $Q$ represents a model or approximation of that process, the KL divergence indicates how well the model $Q$ captures the behavior of the true process $P$. Another interpretation is that $D(P|Q)$ reflects the probability rate of observing a typical trajectory generated by the process $Q$ in the original process $P$. 

\bibitem{R72} R. Blahut, {\it Computation of channel capacity and rate-distortion functions}, IEEE Transactions on Information Theory {\bf 18}, 460 (1972).

\bibitem{S72} S. Arimoto, {\it An algorithm for computing the capacity of arbitrary discrete memoryless channels}, IEEE Transactions on Information Theory {\bf 18}, 14 (1972).

\bibitem{C06} T. M. Cover, {\it Elements of information theory} (Wiley-Interscience, 2006).

\bibitem{S11} S. Shieh, {\it Eigenvalue estimates using the Kolmogorov-Sinai entropy}, Entropy {\bf 13}, 2036 (2011)

\bibitem{AG06} D. Andrieux and P. Gaspard, {\it Fluctuation theorems and the nonequilibrium thermodynamics of molecular motors}, Phys. Rev. E {\bf 74}, 011906 (2006).

\bibitem{FN04} The same result is obtained by using the property (\ref{m.def.alt}), so that $Q$ belongs to the $m$ class of $P$ if $q_1 - q_2 = p_1 -p_2$ since the stationary state is uniform for all parameters. 

\bibitem{A22} D. Andrieux, {\it Fully symmetric nonequilibrium response of stochastic systems}, arXiv:2205.10784 (2022).

\bibitem{A24a} D. Andrieux, {\it Revealing hidden structures and symmetries in nonequilibrium transport}, arXiv:2401.14496 (2024).

\bibitem{FN99} The existence of a potential function $f$ implies that the line integral along any curve $C$ from point $a$ to point $b$ is given by $f_b - f_a$. The physical consequences of this path independence remain to be explored.

\bibitem{AE24} T. Aslyamov and M. Esposito, {\it Nonequilibrium Response for Markov Jump Processes: Exact Results and Tight Bounds}, Phys. Rev. Lett. {\bf 132}, 037101 (2024).

\bibitem{SC12} D. Sivak and G. E. Crooks, {\it Thermodynamic metrics and optimal paths}, Phys. Rev. Lett. {\bf 108}, 190602 (2012).

\bibitem{ZS12} P. R. Zulkowski, D. A. Sivak, G. E. Crooks, and M. R. DeWeese, {\it The geometry of thermodynamic control}, Phys. Rev. E. {\bf 86}, 0141148 (2012).

\bibitem{QEA23} K. N. Quinn, M. C. Abbott, M. K. Transtrum, B. B. Machta and J. P. Sethna, {\it Information geometry for multiparameter models: new perspectives on the origin of simplicity}, Rep. Prog. Phys. {\bf 86}, 035901 (2023).


\bibitem{FN03} This can be illustrated by calculating the symmetrized Kullback-Leibler divergence $D_S (P,Q) = D(P|Q) + D(Q|P)$ between two models within a given equivalence class. It takes the general form $2 D_{S} (P,Q) = (\pmb{A}_P-\pmb{A}_Q)\cdot (\pmb{J}_P-\pmb{J}_Q) + (\pmb{X}_P - \pmb{X}_Q) \cdot (\pmb{Y}_P - \pmb{Y}_Q)$, which reduces to  
$2 D_{S} (P,Q) = (\pmb{A}_P-\pmb{A}_Q) \cdot (\pmb{J}_P-\pmb{J}_Q)$ when both $P$ and $Q$ belong to either an $e$- or $m$-equivalence class. That is, the symmetric variables $\pmb{X}$ and $\pmb{Y}$ do not contribute to $D_S(P,Q)$, which is not the case for two arbitrary dynamics $P$ and $Q$ \cite{A24c}.

\bibitem{C10} G. E. Crooks, {\it On measures of entropy and information}, Tech. Note 009 (2010).

\bibitem{GA11} P. Gaspard and D. Andrieux, {\it Nonlinear transport effects in mass separation by effusion}, J. Stat. Mech. P03024 (2011).

\bibitem{AG08} D. Andrieux and P. Gaspard, {\it Temporal disorder and fluctuation theorem in chemical reactions}, Physical Review E {\bf 77} 031137 (2008).

\bibitem{GG20} J. Gu and P. Gaspard, {\it Counting statistics and microreversibility in stochastic models of transistors}, J. Stat. Mech. P103206 (2020).

\bibitem{HK16} G. E. Henter and W. B. Kleijn, {\it Minimum entropy rate simplification of stochastic processes}, IEEE Trans. on Pattern Analysis and Machine Intelligence {\bf 38}, 2487 (2016). 

\bibitem{FN01} The simplification scheme of Henter and Kleijn requires increasing the system's affinities. This might have interesting implications on model denoising and simplification. 

\end{thebibliography}
\end{document}